\documentclass[11pt,tightenlines,nofootinbib,amsmath,amssymb,aps,prd]{revtex4}
\usepackage{amsfonts,amssymb,amsmath,exscale,bbm}
\usepackage{graphicx}
\usepackage[caption=false]{subfig}

\usepackage{amsmath,amssymb}
\usepackage{amsfonts,amssymb,mathrsfs}

\usepackage[bookmarks=false,pdfstartview=FitH]{hyperref}
\usepackage[all]{hypcap}

\def\be{\begin{equation}}
\def\ee{\end{equation}}

\def\f{\frac}

\def\bra{\langle}
\def\ket{\rangle}
\def\dd{{\rm d}}

\def\vp{\varphi}

\newcommand{\hphi}{\widehat{\varphi}}

\usepackage{color}

\begin{document}
\title{\bf The universe as a quantum gravity condensate}

\author{Daniele Oriti} \email{daniele.oriti@aei.mpg.de}
\affiliation{Max Planck Institute for Gravitational Physics (Albert Einstein Institute),\\
Am M\"uhlenberg 1, 14476 Golm, Germany, EU}

\begin{abstract}
This is an introduction to the approach to the extraction of cosmological dynamics from full quantum gravity based on group field theory condensates. We outline its general perspective, which sees cosmology as the hydrodynamics of the fundamental quantum gravity degrees of freedom, as well as its concrete implementation within the group field theory formalism. We summarise recent work showing the emergence of a bouncing cosmological dynamics from a fundamental group field theory model, and provide a brief but complete survey of other results in the literature. Finally, we discuss open issues and directions for further research.
\end{abstract}

\maketitle

\section{Introduction}
\noindent We work on quantum gravity because we want to answer, among others, two fundamental questions: what is the universe made of? what happens to it in extreme physical situations like the big bang, where General Relativity breaks down? 

\noindent The first has to do with the very definition of a theory of quantum gravity, the identification of the candidate microscopic degrees of freedom of spacetime and geometry and matter (the \lq universe\rq , after the relativistic revolution) and their fundamental dynamics. The second requires modelling the large scale features of the same degrees of freedom {\it from within} the complete theory, under some approximation, and showing that they reproduce observational aspects of the universe as described by semi-classical relativistic cosmological models, while at the same time completing them with a deeper understanding of the regimes beyond their range of applicability. In fact, this last issue is also necessarily part of the first. Only after having a convincing story of how the usual description of spacetime and geometry at large scales arises from the more fundamental quantum gravity one, we can claim to have identified solid candidates for the microscopic building blocks of the universe. This is \lq\lq the problem of emergent spacetime in quantum gravity\rq\rq \cite{emergence}.

\noindent This still the outstanding problem of most quantum gravity approaches, despite decades of successes on various fronts. This is not to say, of course, that there has not been considerable progress also in this respect. 
On the one hand, a number of simplified models have been developed, which, while not yet derived from or embedded into a fundamental theory, are however directly inspired by it. One example is loop quantum cosmology \cite{LQC}, which imports insights from loop quantum gravity \cite{LQG} into minisuperspace quantization of cosmological spacetimes, and, beside a large number of interesting formal developments (including mechanisms for the replacement of the cosmological singularity with a quantum bounce), even suggests phenomenologically testable effects \cite{LQC-pheno}.

\noindent On the other hand, fundamental quantum gravity formalisms have made their first moves towards a derivation of cosmology from first principles, using a variety of strategies and with varying degree of success. For examples of such attempts see \cite{SFcosmo, ReducedLQG, cosmoCDT}. 

\noindent In this contribution, we motivate and review one recent line of research which addresses this issue, in the context of the group field theory formalism for quantum gravity \cite{GFT}, in turn strictly related to loop quantum gravity, tensor models \cite{tensors} and lattice quantum gravity \cite{latticeQG}: {\it group field theory condensate cosmology}. Its promise lies in the results already obtained, of course, but even more in the guiding ideas and in the potential for further developments, which we will try to elucidate.

\section{The group field theory formalism}
\noindent Group field theories (GFTs) \cite{GFT} are quantum field theories on a group manifold, characterized by a peculiar type of (combinatorially) non-local interactions. 

\vspace{0.1cm}

\noindent The basic variable is a (complex) field $\varphi \,:\, G^{\times d} \rightarrow \mathbb{C}$, with $G$ a Lie group and $d$ an integer. The key point is that the domain of the fields should not be interpreted as a spacetime manifold. Rather, GFT models of quantum gravity should be understood as {\it quantum field theories of spacetime}, and spacetime should emerge from them only in some regime. The classical phase space of each \lq quantum\rq of the GFT field, an \lq atom of space\rq, is then $\left( \mathcal{T}^*G\right)^d$, while its corresponding Hilbert space of states is $\mathcal{H} \,=\,L^2\left( G^d\right)$. The fields $\varphi(g_i)=\varphi(g_1,...,g_d)$ and the quantum states can also be written in terms of dual Lie algebra variables, (the \lq momenta\rq of the GFT quanta) via non-commutative Fourier transform \cite{non-comm}, or in terms of irreducible representations of $G$, which label a complete basis of the Hilbert space. For 4d Lorentzian quantum gravity models, and in absence of additional, matter-like degrees of freedom, the relevant group is usually the Lorentz group $SO(3,1)$  ($Spin(4)$ in the Riemannian case) or its rotation subgroup $SU(2)$. Each GFT quantum can be depicted as a topological three-dimensional polyhedron with $d$ faces \cite{GFT-all} (or as a vertex with $d$ outgoing links) labelled by the $d$ arguments of the field. In fact, 3-simplices, i.e. tetrahedra, are the most common choice, corresponding to $d=4$. A guiding principle for model building is the requirement that the same polyhedra are {\it geometric} ones (in the sense of piecewise-flat geometry), at least in a classical limit. This translates in precise \lq geometricity\rq conditions on the fields and their dynamics (the \lq simplicity\rq conditions used in spin foam models \cite{SF}). The group, Lie algebra or representation variables labelling GFT states acquire then the interpretation as discrete connection or metric variables, encoding the geometry of the associated polyhedral structures. 

\vspace{0.1cm}

\noindent The GFT (kinematical) Hilbert space is a Fock space 

$$\mathcal{F}\left( \mathcal{H}\right) \, =\, \bigoplus_{N=0}^{\infty} sym \left\{ \mathcal{H}^{(1)} \otimes \cdots \otimes \mathcal{H}^{(N)}\right\}\qquad \mathcal{H} \,=\,L^2\left( G^{\times d}\right) \qquad , $$ 
where we have {\it assumed} bosonic statistics, and we use ladder field operators 
$\hat{\varphi}(g_{1},g_{2},g_{3},g_{4}) , \quad \hat{\varphi}^\dagger(g_{1},g_{2},g_{3},g_{4})$ \cite{GFT-2nd}. Generic states of this Fock space, build out of a Fock vacuum which corresponds to a total absence of either topological or discrete geometric structures, are arbitrary collections of tetrahedra, including those corresponding (via appropriate restrictions) to connected simplicial complexes. Each combinatorial pattern corresponding to a choice of connectivity among the quanta encodes a specific simplicial topology of space, and the GFT formalism naturally allows for quantum superpositions of the same. Equivalently, GFT states can be associated to (superpositions of) open spin network states, analogous to those of loop quantum gravity \cite{GFT-2nd}. Interesting operators, whose meaning is suggested by the simplicial geometric interpretation of the quantities associated to the GFT quanta, can be constructed in the same 2nd quantized language. In fact, one can construct, as GFT observables, all the 2nd quantized counterparts of the kinematical operators of loop quantum gravity, including operators encoding the quantum dynamics of the theory, like the Hamiltonian constraint, suitably re-expressed as acting on the GFT Fock space of spin networks \cite{GFT-2nd}. 
Simple one-body operators are of the form $\widehat{O} = \int \dd \tilde{g}_i \dd g_i \,\, \hat{\varphi}^\dagger(\tilde{g}_i) \, O(\tilde{g}_i,g_i) \, \hat{\varphi}(g_i)\quad ,$
where $O(\tilde{g}_i,g_i)$ are the matrix elements, in the group representation, of 1st quantized operators in the Hilbert space associated to individual tetrahedra (or spin network vertices). The simplest of such operators is of course the total number operator, which simply counts the number of GFT quanta in a give state:  $\widehat{N} = \int \dd \tilde{g}_i \dd g_i \,\, \hat{\varphi}^\dagger(\tilde{g}_i) \, \hat{\varphi}(g_i)$.

\vspace{0.1cm}

\noindent The dynamics of each GFT model is given in terms of an action, which will comprise a quadratic (free) part and one or more higher order (interaction) parts. The peculiarity of GFTs with respect to ordinary field theories is that the pairing of field arguments in the interactions is {\it combinatorially non-local} in the sense that they are not simply identified to one another across the various fields (so that interactions will take place at the same \lq\lq point\rq\rq on the group manifold), but only according to specific patterns of identifications, whose combinatorics is part of the definition of the model. For the most common models, based on simplicial interactions (associated to 4-simplices), the only interaction terms are five-valent: 
\begin{align} \label{gft-action}
S_\lambda = \int & \dd g_{v_1} \dd g_{v_2}  \bar\vp(g_{v_1}) \vp(g_{v_2}) \, K_2
+ \f{\lambda}{5} \int \left(\prod_{a=1}^5 \dd g_{v_a} \bar\vp(g_{v_a}) \right) \bar{\mathcal{V}}_5
+ \f{\lambda}{5} \int \left(\prod_{a=1}^5 \dd g_{v_a} \vp(g_{v_a}) \right) \mathcal{V}_5,
\end{align}
where the complex-valued functions $K_2 := K_2(g_{v_1},g_{v_2})$ and $\mathcal{V}_5 := \mathcal{V}_5(g_{v_1}, g_{v_2}, g_{v_3}, g_{v_4}, g_{v_5})$.  Each $g_{v_a}$ denotes 4 group elements $g_{{(v_a)}_i}$, and for simplicity of presentation we have included only two of the possible terms (with the same coupling constant $\lambda$), one involving only the fields and one only their complex conjugates. With this choice of combinatorics, the possible interaction processes in the quantum theory, corresponding to GFT Feynman diagrams,  are dual to 4d simplicial complexes of arbitrary topology, sort of discretized spacetimes. These arise in the perturbative expansion of the partition function (and the transition amplitudes) of the given GFT model: 
\begin{equation}
Z\,=\,\int \mathcal{D}\varphi\mathcal{D}\bar\vp\; e^{- \, S_\lambda} \, =\, \sum_{\Gamma}\frac{\lambda^{n_\Gamma}}{s(\Gamma)}\,\mathcal{A}_\Gamma \; \label{GFTpath},
\end{equation}
each weighted by a corresponding Feynman amplitude $\mathcal{A}_\Gamma$, and a symmetry factor $s(\Gamma)$ counting the order of the automorphism group of the diagram $\Gamma$ with $n_\Gamma$ interaction vertices. GFT Feynman amplitudes can be written in several equivalent forms: a) as lattice gauge theories for gravity (when written only in terms of group variables) \cite{mike-carlo1}; b) as lattice gravity path integrals for 1st order (BF-like) gravity actions (when written in terms of Lie algebra variables) \cite{io-aristide}; c) as spin foam models, i.e. sum over histories of spin network states (when written in terms of group representations) \cite{mike-carlo2}. 

\vspace{0.1cm} 

\noindent The last fact makes clear how GFTs provide a complete definition of the covariant formulation of the dynamics of loop gravity spin networks \cite{SF}, via field theory methods, in addition to being a 2nd quantized reformulation of its canonical, operator-based formulation. In turn, the possibility of writing down explicitly GFT amplitudes in terms of simplicial gravity path integrals, one for each simplicial complex summed over in the perturbative GFT expansion, shows that the GFT quantum dynamics combines the ideas of quantum Regge calculus and dynamical triangulations \cite{latticeQG}. To complete the picture of relations between GFT and other quantum gravity formalisms, notice that stripping GFT models of their group-theoretic data (which can be done for instance by reducing the domain $G$ to a finite group or any finite index set) and retaining only their combinatorial structures, one obtains a statistical theory of {\it random tensors} \cite{tensors}. This is another fertile area of recent developments, whose results have fed back on GFT allowing a lot of further progress, especially in the context of GFT renormalization \cite{GFTrenorm}. This is also one context in which the advantages of the GFT formalism, with respect to the other ones it is related to, become manifest: it allows to adapt and use standard QFT techniques and ideas to the solution of open problems in quantum gravity, while remaining fully background independent.

\section{Cosmology as quantum gravity hydrodynamics}
\noindent GFTs paint a picture of the microscopic quantum structure of space and spacetime which is discrete and algebraic in nature. Any geometric interpretation applies only to a small set of quantum states and histories and only in some approximation; and when it does, it anyway corresponds only to piecewise-flat, not smooth geometries. Recovering the latter, with the usual notions of space and time, is a key challenge. In this sense, the fundamental degrees of freedom of the theory can be said to be non-spatiotemporal, and spacetime and geometry are {\it emergent}. So, in which regime of the theory should we look for an effective description in terms of smooth spacetime and geometries? Which class of quantum states and which regime of the dynamics reproduces General Relativity and, in particular, its cosmological sector? 

\vspace{0.1cm}

\noindent The basic intuition, as well as our experience with piece-wise flat geometries, and with the various quantum gravity formalisms related to GFTs, suggest that this regime is one in which a large number of the microscopic degrees of freedom is involved. For modelling specific physical situations, one can in principle use simple approximations only involving a few GFT quanta or very simple GFT interaction processes, but these simple approximations apply in very limited cases and, even then, one should learn to control what happens when more degrees of freedom are included, to have any trust in the drastically simplified truncation initially used. Therefore, we have to explore a new direction, within the theory, that along which more and more degrees of freedom are taken into account, moving from the regime in which only simple spin network graphs or simplicial complexes are used. Notice that this is a {\it distinct} and possibly independent direction of approximation with respect to the quantum-classical direction. 

\vspace{0.1cm}

\noindent The natural tool to move along this direction and to explore the continuum phase diagram of GFT (and spin foam) models is the renormalization group. The goal is to identify one (or more) continuum phase where an effective description of the quantum dynamics in terms of smooth geometry and fields is possible, and that is governed by some modified version of General Relativity. Indeed, GFT renormalization has become a very active direction of research \cite{GFTrenorm}, offering a complementary way to study the renormalization of spin foam models, which can also be tackled from a lattice gauge theory perspective \cite{SFrenorm}. At the non-perturbative level, one interesting outcome are many hints of a condensate phase, emerging in a wide range of GFT models, characterised by a non-zero expectation value of the GFT field operator. This provides some support to the development of GFT condensate cosmology. 

\vspace{0.1cm}

\noindent Before we move to that subject, let us discuss the general problem of understanding cosmology from within a full quantum gravity formalism, and the perspective we adopt.

\vspace{0.1cm}

\noindent If quantum gravity is to be understood, as in the GFT formalism, as a quantum theory of the microscopic building blocks of spacetime, and the gravitational field, just as spacetime itself, are the emergent result of their collective dynamics, cosmology is naturally understood as the result of coarse graining the same microscopic degrees of freedom up to global observables only. 
In other words, it is natural to look for General Relativistic dynamics within the {\it hydrodynamics approximation} of the fundamental theory, and for cosmology (intended here are the simplest case of spatially homogeneous continuum geometries) in the most coarse grained sector of the same effective hydrodynamics. Another way to argue in the same direction is to re-examine the content of the \lq Cosmological Principle\rq from a quantum gravity perspective, as a heuristic guide to the identification of the cosmological sector of the theory (see also \cite{CarloFrancesca}). In usual spacetime language (with manifolds, continuum fields, etc), the principle states that \lq\lq every spatial point is physically equivalent to any other\rq\rq which translates into a condition of homogeneity of spatial geometries. In practice, it means assuming that the universe is in a state such that inhomogeneities can be neglected in comparison with the dynamics of the homogeneous modes, i.e. we can neglect degrees of freedom on length scales shorter than the global scale factor. This is indeed the same spirit of an hydrodynamic approximation, based on neglecting the dynamics of the microscopic degrees of freedom with respect to the global ones or, better, coarse graining them until they are collectively captured by a single statistical distribution over the space of \lq single-particle\rq ~data \cite{BEC}. 
{\it Cosmology should then be the hydrodynamics of quantum gravity} \cite{CosmoQGhydro}.

\vspace{0.1cm}

\noindent A rigorous coarse graining procedure in the full theory, leading to such effective hydrodynamics description, is of course very difficult to define and control. However, we can already guess what the result of any proper construction should be. One expects to obtain a description of {\it the macroscopic universe as a fluid}, whose \lq atoms\rq are the GFT quanta, and whose {\it main collective variable is a sort of \lq density\rq ~function}, plus a \lq velocity\rq ~function. The two variables could be combined in a single complex scalar field. In line with the cosmological interpretation of this level of description, these two collective variables should have a {\it domain isomorphic to the minisuperspace of spatially homogeneous geometries} (or their conjugate space of spatially homogeneous extrinsic curvatures), extended by appropriate matter degrees of freedom. 
Notice that, if the standard procedure leading to the reduced 1-particle density from the coarse graining of generic many-particles states is to be reproduced also in this quantum gravity/cosmology context, also the microscopic quanta of the full theory should have an associated phase space matching the phase space of homogeneous geometries (plus matter fields). 
We should also expect the above collective variables describing the fluid-universe to be governed by {\it non-linear dynamical equations}. 
Observables encoding cosmological quantities, which could in principle be computed from the microscopic quantum gravity theory, should also have an effective expression of a statistical type, involving averages over minisuperspace computed using the collective density function. These observables would also satisfy non-linear dynamical equations.  This general vision is indeed realised in GFT condensate cosmology. 

\vspace{0.1cm}

\noindent Even if the the above description of the universe as a quantum gravity fluid is very general, it may turn out to be realisable or appropriate only in one of the (possibly) many macroscopic phases in which the same fundamental quantum gravity system may organize itself. If we assign a physical significance to the non-spatiotemporal building blocks of the universe, as well as to all their macroscopic phases, and it turns out that geometry and cosmology emerge only in one such phase, then we can say that spacetime and the universe {\it emerge} in quantum gravity through a phase transition from some non-geometric phase. This idea has been suggested in various contexts and dubbed {\it geometrogenesis} \cite{geometrogenesis}. Our GFT condensate cosmology is a tentative realization of it, also in light of the mentioned results on GFT renormalization suggesting this type of phase transition.

\vspace{0.1cm}

\noindent The step from the microscopic quantum description to the effective cosmological one, involving a coarse graining of an infinite dimensional set of quantum interacting degrees of freedom, which are moreover non-spatiotemporal in nature and described in a background independent language, should be expected to be highly non-trivial, if possible at all. The situation should be compared to the analogous problem of extracting, directly from the quantum field theory describing the atoms forming it, the effective hydrodynamic description in some phase of a condensed matter system (without the crucial help of a detailed understanding of its symmetries and conservation laws). 

\vspace{0.1cm}

\noindent There is however one case in which this step is rather straightforward, while remaining non-trivial. This is the very rich case of quantum condensates or superfluids \cite{BEC}. Here, the effective hydrodynamic equations of the condensate can be derived even analytically from the quantum dynamics of the atomic theory, at least in the simplest approximation schemes. Looking at the effective hydrodynamics of GFT condensate states, and checking its possible cosmological interpretation, is therefore both a very interesting and a rather convenient strategy, to test the general picture outlined above, and to try to extract physical consequences from such quantum gravity formalism.       

\section{GFT condensate cosmology}
\noindent Let us now describe the recent work on GFT condensate cosmology, a concrete realization of the perspective just outlined. We first present the general scheme and results that are to a large extent model-independent. Then, we discuss in more detail one specific implementation, which shows its main ideas at work and its promise. We also survey briefly the many other results obtained in this research programme. Finally, we offer some answers to a number of frequently asked questions, and put our framework in relation with similar lines of research in related formalisms.

\subsection{The general scheme}
\noindent The general scheme of GFT condensate cosmology was proposed in \cite{GFTcosmo}. It tackles two issues: 

1) the identification of quantum states in the full theory with a consistent interpretation as continuum cosmological spaces;

2) the extraction, from the fundamental quantum dynamics, of an effective dynamics for this restricted class of states, which can be understood as cosmological evolution.

\vspace{0.1cm}

\noindent The quantum states we look for should have the following features: 

a) they are formed by an infinite (or at least very large) number of GFT quanta, i.e. spin network-type degrees of freedom; 

b) they encode the information corresponding to the phase space of homogeneous geometries in terms of a suitable probability distribution.

The first requirement follows from the desire to have approximately continuum spaces, thus involving a large number of the degrees of freedom that GFTs associate to finite graphs or simplicial complexes. The second restricts the search to cosmological data only, which is of course a very small set of observables, and the underlying quantum nature of the formalism requires that they are encoded in a probabilistic manner only. The combination of the two requirement implies that the same quantum states should involve some implicit or explicit coarse graining. 

The {\it condensate hypothesis} suggests to look for states with a special form: they should endow each fundamental 3-simplex or spin network vertex with the same information, at least in first approximation. If the atoms of space were described by a local QFT, this would basically fix uniquely the form of the states. In GFT, we have the possibility of associating different combinatorial patterns to states consisting of several GFT quanta, even when they are individually in the same configuration, corresponding to the different ways in which we can glue tetrahedra (spin network vertices) along boundary faces (outgoing links). We can also form superpositions of the same combinatorial patterns. The condensate requirement can thus be realised in many ways. 

The simplest realization of the above criteria is given by the state that, while associating the same wave function to each GFT quantum, neglects all the connectivity information. It corresponds to an infinite superposition of states describing disconnected tetrahedra (or open spin network vertices). This is a coherent state of the GFT field operator:
\begin{equation} \label{def-sigma}
|\sigma\ket = e^{-\|\sigma\|^2/2} \, e^{ \int \dd g_i \:
\sigma(g_i) \: \hat{\varphi}^\dagger(g_i)} |0\ket \qquad \hat{\varphi}(g_i) | \sigma \ket = \sigma(g_i) | \sigma \ket
\end{equation}
where $\|\sigma\|^2 = \int \dd g_v \dd\phi \:
|\sigma(g_v, \phi)|^2 = \bra\sigma| \widehat{N} |\sigma\ket$.
This state corresponds to an infinite superposition of infinitely many spin network degrees of freedom (as it becomes transparent by expanding the exponential operator), but they are captured by a single collective function $\sigma$, the {\it condensate wave function}, which is also the expectation value of the GFT field operator in the same state.
For GFT models of 4d quantum gravity, in which appropriate group-theoretic data are used and suitable constraints ensure the geometricity of the tetrahedra, the domain of the condensate wave function $\sigma$ is isomorphic to the space of continuum spatial geometries (or extrinsic curvatures) at a point, that is in turn isomorphic to the minisuperspace of homogeneous (anisotropic) geometries (or corresponding homogeneous extrinsic curvatures) \cite{GFTcosmo, SteffenIso}. Intuitively, this isomorphism can be understood as follows: one can think of embedding the tetrahedra corresponding to each term in the above superposition into some topological 3-manifold, and try to reconstruct a 3-geometry (and conjugate variables) from the data provided by the same state; the fact that all tetrahedra are governed by the same wavefunction implies that the quantum geometric data associated to each point in space will necessarily be the same; this is the intuitive notion of a quantum homogeneous geometry. This simple class of product states corresponds to a Gross-Pitaevskii approximation of any more general condensate state \cite{BEC}, in which entanglement and correlations between the atoms of the condensate are neglected. Indeed, the connectivity information can be understood precisely in terms of entanglement between GFT quanta \cite{GFT-2nd, GoffredoFabioDaniele}.  More general condensate states capturing topological information in an explicit manner, involving large superpositions of connected graphs, and suitable to go beyond the homogeneous approximation as well, have been constructed in \cite{generalised}, and used to model spherical horizons in \cite{horizons}.

\vspace{0.1cm}

This drastic approximation allows however to extract an effective dynamics of such states in a rather straightforward manner, for any given GFT model defining the fundamental dynamics of interest \cite{GFTcosmo}. This can be done in several equivalent ways. One is to truncate the Schwinger-Dyson equations of the given GFT model, as applied to the above condensate state, to the lowest order. In practice this gives the expectation value, in the condensate state, of the operator counterpart of the classical equations of motion of the same GFT model: 

\be \label{class-eom}
\bra \sigma | \frac{\delta S[\hat{\varphi},\hat{\varphi}^\dagger]}{\delta \hat{\varphi}^\dagger(g_i)} |\sigma\ket  = 0.
\ee

The effective dynamics so obtained is given exactly by the classical GFT equations, with the GFT field replaced by the collective condensate wavefunction $\sigma$. From the perspective of the GFT path integral, it amounts to approximate the full quantum dynamics with the saddle point equation, a mean field approximation. This is the GFT counterpart of the Gross-Pitaevskii hydrodynamics of a Bose condensate derived from the microscopic atomic field theory \cite{BEC}: a non-linear equation for the condensate wavefunction, in turn encoding the density and velocity of the (super)fluid. 

\vspace{0.1cm}

\noindent Notice that, just like the albeit simple states we have selected contain an infinite number of spin network or simplicial degrees of freedom, this very simple truncation of the full quantum dynamics is equivalent to the infinite resummation of the \lq tree-level\rq spin foam amplitudes or equivalently to an infinite sum over their dual triangulations weighted by the simplicial path integral defined by the GFT model. This highly non-trivial result is obtained by a straightforward application of 2nd quantized QFT techniques, provided by the GFT formalism. 

From the cosmological perspective, based on the continuum geometric interpretation of the condensate wavefunction $\sigma$, as a complex function on minisuperspace, what we have obtained is the following: the effective hydrodynamics of GFT condensates is a non-linear extension of a quantum cosmology-like dynamics with the condensate wavefunction replacing the usual quantum cosmology wavefunction; for generic anisotropic condensates, the non-linear term is also non-local on the domain of the same wavefunction, i.e. on minisuperspace.
  
For more involved (and realistic) GFT condensate states, the effective hydrodynamics is of course more difficult to derive, and less straightforwardly related to the microscopic GFT dynamics. However, all the above general features remain true, as it is its interpretation, matching what we had anticipated from general considerations.

\subsection{A concrete realization: bouncing cosmologies from the EPRL GFT model}
\noindent We now detail the above scheme in one specific GFT model, and also recast the emergent dynamics in a more physical language, i.e. in terms of cosmological observables. We follow \cite{GFTbouncing}. 

\vspace{0.1cm}

\noindent The starting point is the GFT formulation of the EPRL spin foam model \cite{EPRL,SF} for Lorentzian 4d gravity. Its key features are: the choice of simplicial combinatorics in the interactions (this can be generalised \cite{GFT-all}); the use of GFT fields over $SU(2)^4$; the embedding of these data into covariant, $SL(2,\mathbb{C})$ ones, which is the main content of the \lq geometricity\rq constraints of the model, and that is implemented in the dynamics, i.e. in the specific form of the kernels $K_2^{EPRL}$ and $\mathcal{V}^{EPRL}_5$.

The next step is to extend the set of degrees of freedom of the model by introducing the ones describing a free, massless, minimally coupled real scalar field. The main motivation is to use the same scalar field as a relational clock, an internal time variable with respect to which we can parametrize the evolution of the universe, while maintaining a background independent and diffeomorphism invariant language. This is a standard strategy in quantum gravity and quantum cosmology \cite{LQC, relational}. The guiding principle for the corresponding model building is the same as for the pure gravity case: obtaining a GFT model whose Feynman amplitudes can be written as simplicial gravity path integrals for gravity coupled to the scalar field \cite{YiMingyiDaniele}, with the discrete scalar field located on the vertices dual to the 4-simplices of the simplicial complex. This involves: 

a) extending the domain of the GFT field to $SU(2)^4 \times \mathbb{R}$:  $\hphi(g_i) \rightarrow \hphi(g_i, \phi)$, with the extra variable $\phi$ representing the matter scalar field; 

b) defining appropriate kinetic and interaction kernels, which, due to the discretization chosen and the characterizing symmetries of this type of matter, are generically of the form: 
\be
K_2(g,\tilde{g}; \phi,\tilde{\phi}) = K_2(g,\tilde{g}; (\phi-\tilde{\phi})^2) \qquad 
\mathcal{V}_5(g_{i1},\ldots,g_{i5}; \phi_1, \ldots, \phi_5) = \mathcal{V}_5(\{g\})\int d\phi \prod_{v=1}^{5}\delta(\phi_v - \phi) \nonumber \, ;
\ee

c) a further approximation, justified by the focus on an hydrodynamic regime, of GFT fields that are slowly varying with respect to the scalar field variable, and which translates into a truncation of the kinetic term to the 2nd order in a derivative expansion with respect to the same variable. 

\vspace{0.1cm}

\noindent Starting from the above model, we consider the same type of simple GFT condensate states introduced in the previous section, but now with the extended domain for the GFT fields and for the condensate wavefunction. On this wavefunction, we then impose an additional restriction to purely isotropic configurations. The detailed way to do so is discussed in \cite{GFTbouncing}, and other analogous procedures can be found in the literature \cite{GFTcosmo, SteffenIso}. The result is a condensate wavefunction that depends only on a single spin variable (when expanded in modes), or a single group element in a $U(1)$ subgroup of $SU(2)$, and the real scalar field: $\sigma_j(\phi)$.

This wavefunction can also be written in more standard hydrodynamic form as: 

$\sigma_j(\phi) = \rho_j(\phi)\, e^{i \theta_j(\phi)}$, i.e. in terms of the fluid density and phase functions.

Writing down the mean field equations, i.e. the effective cosmological dynamics, for this special class of condensate wavefunctions, one gets the following equation for each spin component:

\be \label{full-eom}
A_j \partial_\phi^2 \sigma_j(\phi) - B_j \sigma_j(\phi) + w_j \bar\sigma_j(\phi)^4 = 0 \qquad,
\ee

where $A_j$, $B_j$ and $w_j$ are functions of the spin variable only, which encode the microscopic dynamics of the EPRL model and whose explicit expression can be found in \cite{GFTbouncing}. 

At this point, we assume that we are in a regime of the dynamics in which the interaction term, although non-vanishing, is sub-dominant compared to the linear one. This is of course a statement that involves both the interaction kernel $w_j$ and the condensate wavefunction $\sigma$. In this last respect, it is analogous to a dilute gas approximation in ordinary Bose condensates. From the quantum gravity point of view, it is also the regime where the spin foam (or simplicial path integral) expansion (i.e. the perturbative expansion of the given GFT model) is expected to capture relevant features of the quantum dynamics. It is also the regime where the simple approximation adopted for the vacuum state (neglecting entanglement and correlations between GFT quanta) is expected to be valid (even if it may be valid beyond this regime, as it often happens in BECs).

In this regime, one finds two approximately conserved quantities per spin mode: 
\be
E_j = A_j |\partial_\phi \sigma_j(\phi)|^2
- B_j |\sigma_j(\phi)|^2 + 
\frac{2}{5} \mathrm{Re}\left(
w_j \sigma_j(\phi)^5 \right)  \qquad Q_j = -\f{i}{2} \Big[ \bar\sigma_j(\phi) \partial_\phi \sigma_j(\phi) - \sigma_j(\phi) \partial_\phi \bar\sigma_j(\phi) \Big] \nonumber
\ee
and correspond to a symmetry under shifts of the relational time variable and under $U(1)$ transformations of the condensate wavefunction, respectively.

\vspace{0.1cm}

\noindent This completes the definition of the effective cosmological dynamics that is extracted from the microscopic quantum dynamics of the model. However, we can now translates it into a set of equations for cosmological observables. Here is where the introduction of the additional scalar field variable is crucial, since it allows to define {\it within the full theory} a set of {\it relational} observables with a clear physical meaning. 
The observables of interest are all computed from microscopic 2nd quantized GFT operators, as expectation values in the condensate state: 

- the volume of the universe at given value of the relational time: 

$V(\phi) = \sum_j V_j \bar\sigma_j(\phi) \sigma_j(\phi) = \sum_j V_j \rho_j(\phi)^2$, with $V_j = j^{3/2}$ ;

- the momentum of the scalar field at given relational time: $\pi_\phi = \sum_j Q_j$ ; notice that this is a conserved quantity with respect to the relational time, as required by the continuity equation for a massless free scalar field;

- the energy density of the scalar field at given relational time: $\rho = \f{\pi_\phi^2}{2 V^2} = \f{\hbar^2 (\sum_j Q_j)^2}{2 (\sum_j V_j \rho_j^2)^2}$ .

\vspace{0.1cm}

\noindent We can then use the effective hydrodynamic equation for the condensate wavefunction and the two conserved quantities to derive the dynamics of the volume of the universe. We obtain:
\be 
\left( \f{V'}{3 V} \right)^2 = \left( \f{2 \sum_j V_j \, \rho_j \, {\rm sgn}(\rho_j') \sqrt{ E_j - \f{Q_j^2}{\rho_j^2} + m_j^2 \rho_j^2}}{3 \sum_j V_j \rho_j^2} \right)^2,
\ee
and
\be
\f{V''}{V} = \f{2 \sum_j V_j \Big[ E_j + 2 m_j^2 \rho_j^2 \Big]}{\sum_j V_j \rho_j^2} \qquad ,
\ee
where the derivative is with respect to the relational time $\phi$ and $m_j^2 = B_j/A_j$. We call the above dynamical equations the {\it generalised Friedmann equations} obtained from the quantum dynamics of GFT condensates, in the Gross-Pitaevskii approximation. 

The first thing to notice is that these equations have the correct classical limit, when the Hubble rate is small compared to the inverse Planck time, which happens at large enough volumes, i.e. when $\rho_j^2 \gg |E_j| / m_j^2$ and $\rho_j^4 \gg Q_j^2 / m_j^2$. One can check in fact that, if in the same regime $m_j^2 \, \approx\, 3G_N$ , then one recovers the standard Friedmann equations for a flat universe. For models satisfying this condition, the classical limit is recovered then, with an effective Newton constant which is actually a function of the microscopic parameters of the model (as to be expected for an hydrodynamic approximation); beyond this regime, one can still interpret the effective dynamics as a Friedmann one but with a non-Riemannian geometry and an effective Newton constant  which is state-dependent and function of the relational time \cite{MairiAccel}.

Second, the requirement that one has a non-zero energy density for the scalar field must be imposed in order to have a good interpretation in terms of a FLRW spacetime. This in turn implies that at least one of the $Q_j$ must be non-vanishing. The implication of this fact for the dynamics of the universe as described by our generalised Friedmann equations is that the volume remains positive throughout the cosmic evolution, and with a single turning point. In other words, our effective cosmological dynamics implies the existence of a quantum bounce replacing the big bang singularity. It can also be shown that this quantum bounce is followed by a phase of primordial acceleration driven by quantum effects \cite{MairiAccel}.

Last, one can show that, for simple condensate wavefunctions that have only one single spin component $\sigma_{j_o}$, the dynamics is governed by the simple equations:

\be \label{fr-ss1}
\left( \f{V'}{3 V} \right)^2 = \f{4 \pi G}{3} \left( 1 - \f{\rho}{\rho_c} \right) + \f{4 V_{j_o} E_{j_o}}{9 V} , \qquad \f{V''}{V} = 12 \pi G + \f{2 V_{j_o} E_{j_o}}{V} 
\ee
where one recognises immediately the classical terms and the quantum corrections, with $\rho_c = 3 \pi G \hbar^2 / 2 V_{j_o}^2 \sim (3 \pi / 2 j_o^3) \rho_{\rm Pl}$. These are exactly the modified Friedmann equations used in Loop Quantum Cosmology, up to the state dependent terms depending on $E_{j_o}$. These emerge naturally, albeit approximately, then, as a a special case of the GFT condensate hydrodynamics.

\subsection{Survey of recent results}
We have given more details only on one specific realization of the formalism and some key results obtained there, but the last few years have seen a large number of interesting developments. We survey them here, and we refer to \cite{GFTcosmoReview} for a more extensive review (already a bit out of date!). 

We have discussed mainly the case of very simple GFT condensate states in a Gross-Pitaevskii-type approximation, neglecting all the connectivity information of the spin network graphs entering the superposition that would be part of any realistic definition of a non-perturbative continuum vacuum state of the theory. This is obviously a drastic simplification. Dipole condensates have been studied in \cite{GFTcosmo} but reflect a similar simplification. More realistic GFT condensates states, defined by large superpositions including very refined spin network graphs (thus, triangulations), have been constructed in \cite{generalised}. They use techniques from tensor models for controlling the topology encoded by the graphs and for defining the appropriate refinement moves, but rely also heavily on the 2nd quantised formalism and the group-theoretic data of GFTs. This allowed for a straightforward generalisation to spherically symmetric continuum geometries, and for the application to the study of quantum horizons and their entropy \cite{horizons}, but they have yet to be analysed at the dynamical level. The general problem of extracting an effective dynamics from given candidates for the vacuum states of the theory, and of determining reasonable candidates for such states, has been approached using the notion of \lq fidelity\rq in \cite{fidelity}. 

The effective hydrodynamics of GFT condensates can be turned in the form of an effective Hamiltonian constraint equation acting on the condensate wavefunction, from which one can try to read out the relation with gravitational dynamics and the Friedmann equations. This was anticipated at the general level in \cite{effHamilt} as a way to deal with the classical GFT equations, and in the context of GFT condensate cosmology this has been the strategy followed in \cite{GFTcosmo, SteffenIso,Gianluca}, with interesting results. This is however only possible for the free part of the dynamics. The classical geometric content of the GFT hydrodynamics can also be extracted by assuming that the condensate wavefunction is a function of minisuperspace variables peaked on some specific point on the corresponding phase space and extracting the equations for the classical geometric data determining these phase space configurations. This has been explored in a very first attempt at GFT hydrodynamics in \cite{GFThydro}. The assumption on the form of the condensate wavefunction amounts to considering condensates of fundamental simplices or spin network vertices, which are themselves semi-classical with respect to the simplicial geometry, i.e. they are 1st quantized coherent states. Although this choice is made also in other approaches to the extraction of cosmology from full quantum gravity \cite{SFcosmo, ReducedLQG}, it lacks a compelling justification, and it is actually rather unnatural from the perspective we have outlined in this paper. 

Observables that play the role of cosmological conjugate variables of mini-superspace type in the emergent cosmological dynamics of GFT condensates are 2nd quantized, collective quantities from the point of view of the microscopic theory. This has several implications, still to be explored in full. A preliminary analysis has been done in \cite{SteffenDaniele}, and it has highlighted the role of the GFT occupation number in any effective dynamics (for example, the evolution of the occupation number could drive the suppression of  specific terms in the effective cosmological dynamics (e.g. a cosmological constant?)). In particular, the very cosmological dynamics (e.g. the expansion in the volume of the universe) may be driven to a large extent by the growth of the occupation number for given value of the individual quantum geometric data associated to the GFT quanta. This is the basic mechanism at the basis of the so-called \lq lattice refinement\rq approach to loop quantum cosmology, realised now within the full theory.  
This has been confirmed, among a number of other results, by a number of works analysing in more detail the relation between the emergent cosmological dynamics of GFT condensates and the one defined by loop quantum cosmology, including the underlying kinematical level and the definition of cosmological observables of interest \cite{SteffenIso,Steffen2, Gianluca}. 

More recently, the dynamical selection of condensate wavefunctions dominated by a single spin component has been shown to take place for a rather general class of GFT kinetic terms (and still neglecting the effect of interactions), in \cite{SteffenSpin}.

A number of works have instead explored, with a more phenomenological approach, the effect of GFT interactions on the emergent cosmological dynamics \cite{MairiAccel,Mairi2,Mairi3}. This is obviously a crucial issue. Beside confirming the existence of regimes in which such interactions are subdominant, reproducing the results discussed in the previous section, these analyses have also shown that the effects of GFT interactions can be very interesting. In specific ranges for the GFT coupling constants, and with simple interactions of order four and six added to the effective relational dynamics derived in the previous section, one sees a realization of a cyclic universe model, wth a long lasting accelerated phase right after the cosmological bounce, sustained for a large enough number of e-folds to be a sort of \lq quantum geometric\rq inflationary phase (so without any inflaton field or specific inflaton potentials), and then, after the expansion to macroscopic scales, a re-collapse towards a new bounce. This is reported in \cite{Mairi2}.
Another possible effect of the GFT interactions is to drive solutions of the emergent cosmological dynamics to regimes in which the average occupation number diverges, thus suggesting a breakdown of the Fock representation, and the need for a new inequivalent GFT representation to describe the quantum dynamics with strong interactions. This is consistent with a geometrogenesis scenario. The same GFT interactions can also produce dynamically the dominance of single-spin configurations at late (relational) times that was see in the free models of \cite{SteffenSpin}. These results have been obtained in \cite{Mairi3}. A further analysis of the role of the GFT interactions in the emergent cosmological dynamics has been done in \cite{Mairi4}, where one can find also a first analysis of the role of anisotropies in the emergent cosmological dynamics, and interesting hints of a dynamical isotropization of the universe that takes place very quickly in relational time.

Some first steps have also been taken in the extension of GFT condensate cosmology to the study of cosmological perturbations, restricted to homogeneous metric perturbations \cite{SteffenPert1}, and to inhomogeneous geometries reconstructed from the statistical distribution over homogeneous geometries provided by the condensate density \cite{Steffen2}.

\subsection{Discussion: answers to frequently asked questions and relation with other approaches}
\noindent We clarify here our perspective on a number of criticisms that have been raised to this approach and to some results described above. 

A first objection is against the use of the simple condensate states corresponding to the Gross-Pitaevskii approximation to the condensate hydrodynamics, and it stems from the true fact that such states cannot encode any topological information about quantum space, and they can only encode in a limited way other information about spatial geometry such as curvature (which also relies on connectivity information, e.g. in the definition of the gravitational holonomies). Given this, how can these states be used to represent realistic quantum spaces? The answer is the following. First, even when such simple states are plugged in the quantum dynamics, this does produce states associated to connected graphs, since they are naturally generated by the GFT interactions.  Second, using such states only means assuming that the relevant physical states for describing cosmology are {\it approximated} in some regime by these simple condensate states, i.e. that the information encoded in the connectivity can be {\it approximately} neglected in the choice of reference quantum states at least when focusing on limited aspects of the quantum dynamics. One expects that the dynamics of homogeneous gravitational fields (which is in many ways \lq ultra-local\rq) is one of them. There is no implication that any realistic state of the theory, solving the full quantum dynamics, will be of such simple type. This is all the more true when one notices that spin network connectivity encodes the entanglement between GFT quanta \cite{GoffredoFabioDaniele}, since one expects physical states of interacting quantum systems to be highly entangled. One example is Bose condensates themselves \cite{BEC}. But the same example also teaches us that a Gross-Pitaevskii approximation can capture much useful information about the macroscopic properties of the system, at least in some regime. This is exactly what we aim to obtain, from our quantum gravity condensates.

The second objection reflects the worry that the approximation in which one neglects GFT interactions, often used in the literature, cannot give sensible answers, because in spin foam models one usually encodes most of the information about the quantum dynamics in the spin foam vertex amplitudes, corresponding to the GFT interaction kernels, and this is also where simplicity constraints are usually implemented. This objection can be countered at several levels. First, the simplicity constraints do not encode, per se, any dynamical information as far as gravity is concerned. They are rather a precondition for interpreting the GFT quanta and their dynamics (or the underlying spin foam amplitudes) in geometric terms, and so they are implemented the very moment we use $SU(2)$ data while giving them a geometric meaning. Second, it is possible to include the simplicity constraints and most dynamical information in the kinetic term as well as in the potential in the GFT action, by appropriate field redefinitions. This ambiguity is present also at the spin foam level, where dynamical information can be shifted from the vertex amplitude to the edge amplitude. This is why it is convenient to work with more general forms of kinetic and interaction kernels, rather than specific choices, whenever possible and provided important defining features are implemented. 
Third, in this framework, the interaction term {\it is present}, and so are the simplicity constraints; only, we restrict our attention to the regime where the contribution of the interaction term to the equations of motion is subdominant, to check if the dominant contribution contains already enough cosmological dynamics. The overall scaling of the volume with respect to the relational time field (the whole content of the Friedmann equations) may well be such a simple sector of the dynamics to be already captured in this approximation. 

A related worry is that the mesoscopic regime in which the GFT kinetic term dominates over the GFT interactions, while at the same time the volume, and thus the condensate wavefunction, is large enough to have a good semi-classical approximation, may not exist. A full answer should be of course quantitative, and the phenomenological analyses of interacting GFT models mentioned above are important steps in this direction. At the qualitative level, we mention the following. First of all, the mere existence in a given GFT model of such mesoscopic regime is guaranteed simply because the interactions are weighted by coupling constants which can always be tuned to be small enough, that, when the solution of the free GFT equations is semi-classical, the GFT interactions are subdominant. They are also expected to be small on general grounds, and this is also supported by the renormalization group analysis of simpler GFT models \cite{GFTrenorm}. The real issue is how long (in relational time) this regime lasts.  Once more, this depends on the specifics of the GFT interaction kernels, on their relative amplitudes with respect to the kinetic terms, and on the value of the coupling constants, and has to be studied in the context of specific solutions of the effective cosmological dynamics extracted from specific GFT models. As is the case in LQC, therefore, the requirement of a long enough semi-classical cosmological regime constrains the details of model building in the fundamental theory (i.e., the ambiguities in the spin foam amplitudes and also in the choice of the appropriate GFT action). Both will then be further constrained by renormalization analysis and other formal conditions on the models (e.g. stability against fluctuations).

In most other approaches to the extraction of cosmology from full quantum gravity, the quantum states being used to describe cosmological spacetimes are spin network states with a specific graph dependence and peaked on classical geometries discretised on the lattice dual to the spin network graphs, and both the choice of lattice and the choice of discrete geometry to peak on are selected to match the homogeneity of classical cosmological geometries. This is not what we do in GFT condensate cosmology. The only restriction on microscopic quantum states is the condensate hypothesis, which requires a single wavefunction to capture the whole content of the states. This is also the only sense in which these microscopic states are \lq homogeneous\rq. This is enough to specify the whole structure of simple condensate states in the Gross-Pitaevskii approximation, but not for more involved condensate states  \cite{generalised}. In the latter case, the microscopic states entering the superposition are not themselves homogeneous, for example they do not live on regular graphs. The coarse grained nature of the cosmological data becomes also much more evident. This is consistent with cosmology being understood as quantum gravity hydrodynamics, rather than the result of any symmetry reduction at the fundamental level. It also reflect the distinction between continuum and semi-classical approximation, which we have also emphasized. The continuum approximation is the crucial one that should lead us from the microscopic quantum gravity description to the macroscopic, geometric one, and the result of taking it {\it may depend crucially on the quantum properties} of the microscopic degrees of freedom, which should therefore be maintained when attempting to extract effective macroscopic physics. This is exactly what happens in standard Bose condensates, where the very macroscopic properties of the system in the hydrodynamic approximation are the result of the {\it quantum} nature of the underlying atoms. The GFT condensate wavefunction can be chosen to be \lq semi-classical\rq from the point of view of minisuperspace data, but whether this is useful should only be determined once the effective macroscopic dynamics has been extracted. The semi-classical limit should be a particular regime of the emergent quantum dynamics, rather than be an ab initio constraint on the form of the quantum states.   

We have presented the framework of GFT condensate cosmology motivating it also with the idea of a \lq geometrogenesis\rq phase transition of condensation type. We have also suggested the possibility that such cosmological phase transition could replace the big bang singularity in the fundamental quantum gravity theory. The doubt can arise, therefore, if the existence of such phase transition is actually {\it needed} for GFT condensate states to be useful non-perturbative states for physical purposes, and for GFT condensate cosmology to be a realistic scenario for the universe. This raises a legitimate worry since no evidence exists yet for such transition in realistic 4d gravity models. The answer is negative: the condensate hypothesis for the quantum states used to extract cosmology from the full theory does not {\it rely on} the existence of a condensate phase transition. GFT condensate states may well be used within the Fock representation used to define the perturbative regime and are simply candidates for physically interesting quantum states. Their actual physical interest depends on their effective dynamics, i.e. from their suitability to describe our universe. Whether or not they also identify a different macroscopic phase of the GFT system with respect to simple spin network states, this will not affect their physical relevance. A similar consideration applies to the general idea of cosmology arising in the hydrodynamic approximation of full quantum gravity. This may well be true, whether or not the appropriate quantum states to be used to describe our universe are GFT condensates, which only represent a particularly convenient case to study, since they allow a straightforward extraction of the corresponding hydrodynamic equations.

The perspective on cosmology as the hydrodynamics of quantum gravity also implies a certain attitude towards another recurrent worry: how many of the microscopic details of the quantum dynamics, as encoded in a specific GFT model, are crucial for the emergence of the correct cosmological dynamics? and which ones? conversely, how many features of the effective cosmological dynamics are instead universal? As a general answer, inspired by what happens in condensed matter systems, we should expect that many of the microscopic details that differentiate between GFT models (e.g. specific way of imposing simplicity constraints, specific weights associated to face or edge amplitudes in spin foam models, choice of quantization maps, etc) will not modify drastically the continuum physics extracted from them. The coarse graining procedure may erase them. On the other hand, some structural aspects of the microscopic models can produce drastic differences even at macroscopic level. Think for example at the difference between bosonic and fermionic condensed matter systems. In the end, we will be able to determine what is model-dependent  and what is universal only once we will have a better control on the reconstruction of cosmology from microphysics, on the phase diagram of GFT models, on their symmetries \cite{GFTsymm} (which may largely determine their hydrodynamics) and on the microscopic origin of cosmological phenomena. 

\

\noindent We now compare briefly GFT condensate cosmology with other approaches to the problem of extracting cosmology from quantum gravity.  

The view on cosmology realised via GFT condensates is very similar to the one advocated in \cite{MartinQC}. This is based on the idea that quantum cosmology should be understood as a description of a \lq local patch\rq ~of the universe, small enough to be approximated as homogeneous, and the building block of a multi-patch dynamics that only can capture correctly cosmological physics. Rather than symmetry reduction, the theoretical guiding principle should then be the analogy with condensed matter theory and the physics of many-body quantum systems. This is exactly our picture of \lq cosmology as quantum gravity hydrodynamics\rq. The issue is: what is a \lq patch\rq in our GFT description of quantum gravity. In GFT condensate cosmology, the basic building block to which the collective wave function is associated is directly an individual GFT quantum; this is where the hypothesis of condensation determines the form of GFT hydrodynamics. More generally, a large number of GFT quanta may need to be coarse grained to obtain an elementary piece of continuum space that could serve as the starting point for GFT hydrodynamics, thus a GFT patch. Beside the general philosophy, the perspective in \cite{MartinQC} has been implemented in a non-linear generalization of loop quantum cosmology \cite{nonlinearLQC}, constructed in analogy with Bose condensates. Its equations are in fact very close to the classical GFT equations used to describe cosmological evolution in GFT condensate cosmology (and also to those of the symmetry-reduced models proposed in \cite{GFC}).

GFT condensate cosmology can be seen also as an attempt to extract cosmology from spin foam models, since GFT provides one possible definition of a complete spin foam dynamics. Thus it is natural to compare it with so-called \lq spin foam cosmology\rq \cite{SFcosmo}. This strategy amounts to the straightforward computation of spin foam transition amplitudes for boundary states that admit an interpretation in terms of homogeneous universes, and their re-writing in terms of an effective Hamiltonian constraint equation to be compared with the one of homogeneous GR. Beside technical aspects (see for example \cite{Frank} for a critical analysis), the main differences stem from a change in perspective. While in GFT condensate cosmology one tries to extract an effective hydrodynamic description for the collective observables describing a large number of microscopic spin network degrees of freedom, here the focus is on the spin network degrees of freedom themselves and their fundamental quantum dynamics. This also implies assigning a continuum interpretation directly to these degrees of freedom, by using the corresponding simplicial geometries to describe the physical universe. Moreover, a semiclassical set of states (from the point of view of such simplicial geometries) is used. In themselves, none of these points is in contradiction with GFT condensate cosmology, which indeed agrees on the physical nature of spin network data. However, from the point of view of GFT condensate cosmology, the inevitable approximations used in \cite{SFcosmo} should be expected to be of very limited physical validity, and do not allow to bypass the need to extract a more comprehensive hydrodynamic description for the same spin network degrees of freedom. Moreover, from this hydrodynamic perspective, the very use of the spin foam expansion to study continuum physics is questionable, as it would be the use of the Feynman expansion (around the trivial vacuum state) of an atomic field theory underlying a condensed matter system to study its macroscopic features. One can obtain the basic setting of \cite{SFcosmo} within GFT condensate cosmology in the following way: a) consider a dipole GFT condensate \cite{GFTcosmo}, and truncate it to its lowest order in the number of dipoles (one); b) insert it in the GFT quantum dynamics corresponding to any given GFT (or spin foam) model, generalised to arbitrary 2-complexes (i.e. beyond simplicial geometric ones) \cite{GFT-all}; c) expand this quantum dynamics in perturbative Feynman diagrams and truncate this expansion at the lowest non-trivial perturbative order. GFT condensate cosmology does not suggest that anything is wrong with any of these steps and simplifications, but it suggests that they capture interesting continuum gravitational physics only in very exceptional cases and that other methods and approximations should be expected to be more convenient. Of course, only detailed calculations of physical aspects of our universe can tell to what extent this is the case.

In the canonical LQG framework, a series of works \cite{ReducedLQG} has developed another interesting strategy for extracting cosmology from the full quantum gravity theory. The aim is similar, and in fact, a similar operator-based approach can be pursued also within GFT condensate cosmology. The main difference is the focus on a quantum counterpart of the classical symmetry reduction to homogeneity, rather than on the coarse graining to homogeneous observables capturing the collective features of microscopic configurations which are not individually homogeneous in any strict sense. From this different focus, several technical differences follow: the restriction to hypercubic lattices in \cite{ReducedLQG}; the attempt to solve the Hamiltonian constraint equations exactly rather than an effective, hydrodynamic-type version of the same; the difficulty in incorporating large superpositions of spin network states and dynamical lattices; the use of semiclassical coherent states of spin networks. This being said, a number of similarities can be found also at the technical level between the class of states used in \cite{ReducedLQG} and GFT condensates, which encourage to have a closer look at the detailed relation between these two approaches to the same problem.
 
 Finally, we mention that an effective minisuperspace dynamics has been extracted also from the quantum gravity formalism of causal dynamical triangulations \cite{cosmoCDT}. The general strategy adopted there seems quite similar to the one of GFT condensate cosmology: focus on the computation of coarse grained observables like the total volume of the universe in the full quantum dynamics, without any symmetry reduction in the underlying microscopic configurations, with the effective dynamics of such observables reflecting the quantum properties of the underlying theory (no semiclassical approximation taken beforehand) and the continuum limit involving a superposition of a large number of microscopic degrees of freedom. It is difficult, however, to compare the two strategies in detail, because GFT condensate cosmology deals with the same quantum properties and large superpositions, in an approximate manner, via field-theoretic tools applied to the GFT that generates the sum over triangulations, while in causal dynamical triangulations the same sum is dealt with directly by numerical methods. Still, one feels encouraged to explore further similarities and differences between the two formalisms.

\section{The way ahead}
\noindent To conclude, we summarise some open issues and research directions within GFT cosmology. 

At the more mathematical level, the whole programme of GFT condensate cosmology will benefit from any advance in our understanding of solutions to the full quantum dynamics of interesting GFT models, i.e. in the characterization of realistic vacuum states of the full theory, in particular those which can admit a cosmological interpretation.
Understanding the full quantum dynamics also means gaining a better control over the phase diagram of interesting GFT models of 4d gravity; the goal would of course be to accumulate evidence of a condensate phase of the theory.
We should achieve a better control over the differences produced at the effective cosmological level by different microscopic GFT models, especially in the Lorentzian case. We need to construct appropriate observables encoding spatial curvature and topological properties of the universe, and to extract their contribution to the emergent cosmological dynamics. The same is true for observables capturing cosmological anisotropies. 
Related to this, while more interesting condensates states have been constructed, we have yet to derive their effective hydrodynamic equations, and this is important exactly because such states allow for a richer encoding of geometric and topological properties of the universe. 
Even for simple condensate states, however, in order to give more solidity to the whole approach, it is mandatory to improve our control over the effects of GFT interactions, to develop techniques to solve the corresponding non-linear GFT equations, and to perform detailed analyses of the stability of any solution against fluctuations.

At the physical level, one crucial issue has to do with the fate of cosmological singularities. GFT condensate cosmology is motivated by the possibility of a cosmological phase transition replacing the standard big bang scenario. On the other hand, we have seen that one can obtain a bouncing cosmology from the hydrodynamics of GFT condensates, with a quantum bounce replacing the big bang. A solid realization of the latter scenario in a full quantum gravity formalism would be of course tantalizing. GFT condensate cosmology seems to provide it, but this is true only as long as the hydrodynamic approximation maintains it validity in the regime corresponding to the quantum bounce. This has to be studied in detail. In both scenarios one needs to extract phenomenological signatures of the regime of the theory that replaces the big bang. For a quantum bounce, one can build on considerable experience obtained in LQC \cite{LQC-pheno}, while for cosmological phase transitions there are fewer, if still inspiring, references \cite{CosmoPhasePhen}. 
For doing so, it will be useful to recast the emergent cosmological dynamics extracted from GFT condensates in the form of some effective modified gravitational dynamics, to compare it with existing cosmological scenarios for the very early universe: the inflationary scenario, matter bounce scenarios \cite{matterbounce}, or emergent universe scenarios \cite{emergentUniv}.
In this effective gravitational dynamics, one should identify clearly the quantum gravity origin of spatial curvature terms, topological contributions and of any term interpretable as an effective cosmological constant. Special attention should be payed to the dynamics of anisotropies, looking for a general quantum gravity mechanism driving the isotropization of the universe during its evolution.
One expects the effective gravitational dynamics to contain a number of quantum gravity corrections to the standard relativistic one: corrections to the Friedmann equations simply due to the form of the microscopic quantum dynamics; effects coming from the use of holonomies rather than connection variables, like in LQC, and effects due to the non-commutativity of flux variables encoding metric degrees of freedom; macroscopic signatures of the underlying discreteness of quantum geometry (discrete spectra for geometric operators and lattice-like structure of the fundamental quantum gravity excitations); effects depending on a depletion factor and on the backreaction of degrees of freedom outside the condensate state (e.g. fluctuations over it). 
The real test of the physical significance of GFT condensate cosmology will be its ability to provide a fundamental understanding of cosmological perturbations. This is also where its phenomenological consequences should be found. 
This is of course a technical challenge, as it involves the analysis of fluctuations over condensate states within the full theory. But the biggest hurdle is not technical, as the calculations involved in deriving an effective dynamics for such GFT perturbations are rather straightforward and not too different from standard ones in condensed matter theory. Both the definition of GFT perturbations and their effective dynamics will still be expressed in the fully background independent language of the fundamental theory, purely algebraic and combinatorial, and the real issue is to translate them in a more customary language of fields over the spacetime geometry, encoded in the given GFT condensate state around which one is expanding. This amounts to a derivation of an effective field theory framework for geometry and matter from a full quantum gravity formalism, obviously a non-trivial task. One strategy to meet this challenge is to extend the relational approach used in \cite{GFTbouncing}, by introducing additional degrees of freedom that can be used as rods and clocks to identify spacetime points in truly physical terms, and to parametrise the kinematics and dynamics of GFT perturbations over condensate states \cite{DanieleSteffen}. A first step towards such complete treatment is the implementation within GFT condensate cosmology of the \lq separate universe\rq approach to cosmological perturbations, valid for long wavelengths, already adopted in the context of loop quantum cosmology \cite{EdSeparateLQC}. This requires moving from simple condensate states to the GFT analogue of multi-condensate states, with each condensate component describing an homogeneous patch in an otherwise inhomogeneous universe \cite{EdFlorianDaniele}. The dynamics of each patch is expected to be approximately governed by the modified Friedmann dynamics discussed above, and the dynamics of inhomogeneities can be computed  by comparing the evolution of different patches. Less ambitious still, one can formulate in the context of GFT condensate cosmology the \lq dressed meric\rq approach used in LQC \cite{dressedLQC}, in which one assumes that the effective field theory of cosmological perturbations remains the standard one, but the the evolution takes place on an metric background solving the quantum gravity-modified Friedmann equations. This strategy has the shortcoming of a not too solid foundation, but also the advantage of being straightforward and of giving immediately indications of possible phenomenological consequences of the theory. 

The conceptual aspects of the approach to cosmology exemplified by GFT condensate cosmology are not less interesting. One set of issues has to do with the general problem of spacetime emergence in quantum gravity \cite{emergence,ChrisNick}, for which GFT condensate cosmology provides a tentative (partial) solution. There are then conceptual issues around the notion of geometrogenesis and its physical interpretation as a dynamical process, a physical event replacing the big bang singularity. This idea raises a host of philosophical questions, which have also to do with the possibility of attributing a dynamical characterization to the renormalization group flow in theory space, and to renormalization group trajectories in the continuum phase diagram of GFT models (or other quantum gravity models). This is truly unexplored territory in the philosophy of physics literature, to the best of our knowledge. A number of conceptual issues have instead to do with the general idea of cosmology as the hydrodynamics of quantum gravity. Not only this represents a departure from the more traditional, albeit problematic, quantum cosmology framework, but it also comes with a new bundle of questions. These concern for example the collective nature of cosmological observables, the significance of the \lq cosmological wavefunction\rq appearing as the fundamental dynamical variable in GFT condensate cosmology, and especially the role and consequences of the inevitable non-linearities in the associated dynamics  \cite{CosmoQGhydro}. 
They also represent unexplored territory.

\

\noindent It should be clear, therefore, that like all exciting research directions, GFT condensate cosmology is attractive because of the new questions it raises  as much as for the issues it promises to solve.

\end{document}